\documentclass[5p]{elsarticle}
\usepackage{amsmath}
\usepackage{graphicx}
\usepackage{url}
\biboptions{sort&compress}
\journal{Physics Letters B}

\begin{document}
\begin{frontmatter}

\title{Low-energy electric dipole response in $^{120}$Sn}

\author[ikp]{A.M.~Krumbholz}\ead{amkrumbholz@ikp.tu-darmstadt.de}
\author[ikp]{P.~von~Neumann-Cosel\corref{cor1}}\ead{vnc@ikp.tu-darmstadt.de}
\author[rcnp,dae]{T.~Hashimoto}
\author[rcnp]{A.~Tamii}
\author[rcnp]{T.~Adachi}
\author[com]{C.A.~Bertulani}
\author[rcnp]{H.~Fujita}
\author[osaka]{Y.~Fujita}
\author[ist]{E.~Ganioglu}
\author[rcnp]{K.~Hatanaka}
\author[rcnp]{C.~Iwamoto}
\author[rcnp]{T.~Kawabata}
\author[rcnp,inst]{N.T.~Khai}
\author[ikp]{A.~Krugmann}
\author[ikp]{D. Martin}
\author[chiba]{H.~Matsubara}
\author[itl]{R.~Neveling}
\author[rcnp]{H.~Okamura\fnref{fn1}}
\author[rcnp]{H.J.~Ong}
\author[ikp]{I.~Poltoratska}
\author[ikp]{V.Yu.~Ponomarev}
\author[ikp]{A.~Richter}
\author[miy]{H.~Sakaguchi}
\author[toh]{Y.~Shimbara}
\author[riken]{Y.~Shimizu}
\author[ikp]{J.~Simonis}
\author[itl]{F.D.~Smit}
\author[ist]{G.~Susoy}
\author[mun]{J.H.~Thies}
\author[rcnp]{T.~Suzuki}
\author[rcnp]{M.~Yosoi}
\author[riken]{J.~Zenihiro}

\cortext[cor1]{Corresponding author}
\fntext[fn1]{Deceased.}

\address[ikp]{Institut f\"ur Kernphysik, Technische Universit\"at Darmstadt, D-64289, Darmstadt, Germany}
\address[rcnp]{Research Center for Nuclear Physics, Osaka University, Ibaraki, Osaka 567-0047, Japan}
\address[dae]{Rare Isotope Project, Institute for Basic Science, 70 Yuseong-daero, 1689-gil, Yuseong-gu, Daejeon, Korea}
\address[com]{Department of Physics, Texas A\&M University-Commerce, Commerce, Texas 75429, USA}
\address[osaka]{Department of Physics, Osaka University, Toyonaka, Osaka 560-0043, Japan}
\address[ist]{Physics Department, Faculty of Science, Istanbul University, 34459 Vezneciler, Istanbul, Turkey}
\address[inst]{Institute for Nuclear Science and Technology, 179 Hoang Quoc Viet, Hanoi, Vietnam}
\address[chiba]{National Institute of Radiological Sciences, Chiba 263-8555, Japan}
\address[itl]{iThemba LABS, PO Box 722, Somerset West 7129, South Africa}
\address[miy]{Department of Physics, Miyazaki University, Miyazaki 889-2192, Japan}
\address[toh]{Cyclotron and Radioisotope Center, Tohoku University, Sendai 980-8578, Japan}
\address[riken]{RIKEN, Nishina Center, Wako, Saitama 351-0198, Japan}
\address[mun]{Institut f\"ur Kernphysik, Westf\"alische Wilhelms-Universit\"at M\"unster, D-48149 M\"unster, Germany}

\begin{abstract}
The electric dipole strength distribution in $^{120}$Sn has been extracted from proton inelastic scattering experiments at $E_{\rm{p}} = 295$~MeV and at forward angles including $0^\circ$. 
Below neutron threshold it differs from the results of  a $^{120}$Sn$(\gamma,\gamma^\prime)$ experiment and peaks at an excitation energy of 8.3 MeV. 
The total strength corresponds to 2.3(2)\% of the energy-weighted sum rule and is more than three times larger than what is observed with the $(\gamma,\gamma^\prime)$ reaction.
This implies a strong fragmentation of the E1 strength and/or small ground state branching ratios of the excited $1^-$states.    
\end{abstract}


\end{frontmatter}

\section{Introduction}\label{intro}

The low-energy electric dipole strength in neutron-rich nuclei, commonly termed Pygmy Dipole Resonance (PDR), is currently a topic of great interest \cite{sav13}. 
It occurs at energies well below the isovector Giant Dipole Resonance (GDR) and exhausts a considerable fraction (up to about 10\%) of the total E1 strength in nuclei with a large neutron-to-proton ratio  \cite{adr05,kli07,wie09,ros13}. 
The properties of the mode are claimed to provide insight into the formation of a neutron skin \cite{pie06,kli07,tso08,pie11,ina11}, although this is questioned \cite{rei10}. 
It may also constrain the density dependence of the symmetry energy \cite{kli07,car10,fat12,tsa12}. 
Thus, investigations of the PDR will be an important topic at future rare isotope beam facilities. 
Furthermore, dipole strength in the vicinity of the neutron threshold may lead to significant changes of neutron-capture rates in the astrophysical $r$-process \cite{gor04,lit09,dao12}.

Originally considered to be a single-particle effect \cite{lan71}, many microscopic models nowadays favor an explanation of the PDR as an oscillation of a neutron skin - emerging with an increasing $N/Z$ ratio - against an approximately isospin-saturated core. 
This conclusion is based on the analysis of theoretical transition densities which differ significantly from those in the GDR region.
However, at least for stable nuclei with a moderate neutron excess this question is far from being settled, see e.g.\ the recent work of Ref.~\cite{pap14}. 
Quantitative predictions of the centroid energy and strength of the PDR as well as the corresponding collectivity  as a function of neutron excess differ considerably.
This is due partly to the properties of the underlying mean-field description (e.g., Skyrme-type or relativistic models) but also results partly from the difficulty to separate clearly the location of PDR and GDR.
E1 strength distributions at low excitation energies are also strongly modified when complex configurations beyond the 1 particle - 1 hole (1p1h) level are included in the models (see e.g.\ Refs.~\cite{rye02,ton10,lit10}). 

Data on the low-energy E1 strength in very neutron-rich heavy nuclei are scarce \cite{adr05,kli07,wie09,ros13}.
Although the PDR strength is much weaker in stable nuclei, detailed spectroscopy with different isovector \cite{kne96,tam11} and isoscalar \cite{pol92,end10,pel14} probes provides important insight into a possible interpretation of the mode as a neutron-skin oscillation, the interplay of collectivity and single-particle degrees of freedom and its isospin nature \cite{paa09,roc12,vre12,yuk12,lan14}.
Extensive studies have been performed in stable even-mass nuclides utilizing the $(\gamma,\gamma')$ reaction, in particular at shell closures.
However, the connection of these results to the PDR in nuclei with very large $N/Z$ ratios is not clear \cite{sav13,paa07}.

$(\gamma,\gamma')$ experiments are selective towards ground-state (g.s.) transitions because the experimental cross sections are proportional to $\Gamma_f \Gamma_0/\Gamma$ and the experimental background limits the analysis to $\Gamma_f = \Gamma_0$ in excitation and decay energy regions of high level density.
Here $\Gamma_0$ and $\Gamma_f$ denote the partial widths to the g.s.\ and any final state $f$, respectively, and $\Gamma$ is the total width.
Possible branching ratios to excited states are often neglected, but statistical model calculations of the branching ratios suggest potentially large corrections of the deduced E1 strength \cite{rus08}.
This uncertainty has an impact on the determination of the E1 polarizability, which has been established as a measure of the neutron skin and the density dependence of the symmetry energy \cite{rei10,pie12}.

In a benchmark experiment on the doubly magic nucleus $^{208}$Pb \cite{tam11,pol12}, relativistic Coulomb excitation in proton scattering at energies of a few hundred MeV and very forward angles has been established as a new promising approach to study the complete E1 strength in nuclei \cite{tam09,nev11}.
The method avoids the above discussed problem of decay experiments and allows consistent measurements of the E1 strength below and above neutron threshold, thereby providing precise values of the polarizability \cite{tam11,tam14}.

The present letter discusses such an experiment for the semimagic nucleus $^{120}$Sn.
The tin isotope chain is of special interest as it allows a systematic study of the properties of the PDR in nuclei with similar structure features but varying neutron excess (see e.g.\ Refs.~\cite{sav13,oze14} and references therein). 
Data from $(\gamma,\gamma')$ measurements are available for $^{112,116,120,124}$Sn~\cite{gov98,oze14}. 
Comparing the extracted E1 strength distributions, $^{120}$Sn shows a stronger fragmentation pattern than the other isotopes and a local minimum of the integrated strength, while the results in the other three isotopes would be consistent with a correlation between PDR strength and neutron excess \cite{oze14}. 
The new experimental results presented here show that the major part of the E1 strength distribution in $^{120}$Sn up to neutron threshold is missed in the $(\gamma,\gamma')$ experiment indepedent from possible corrections due to decays to excited states.

\section{Experiment}
\label{sec:experiment}

The experiments were performed at the Grand Raiden spectrometer of the Research Center of Nuclear Physics in Osaka using a 295~MeV polarized proton beam at spectrometer angles $\Theta = 0^\circ$, $2.5^\circ$ and $4^\circ$. 
The detector setup and the principles of the raw-data analysis are described in Ref.~\cite{tam09}.
Details of the conditions and the analysis of the present measurements can be found in Ref.~\cite{kru14}.
The present work focuses on the information from a multipole decomposition of the cross sections; polarization transfer results will be discussed elsewhere \cite{has15}.

\begin{figure}[t]
\centering
\includegraphics[width=\columnwidth]{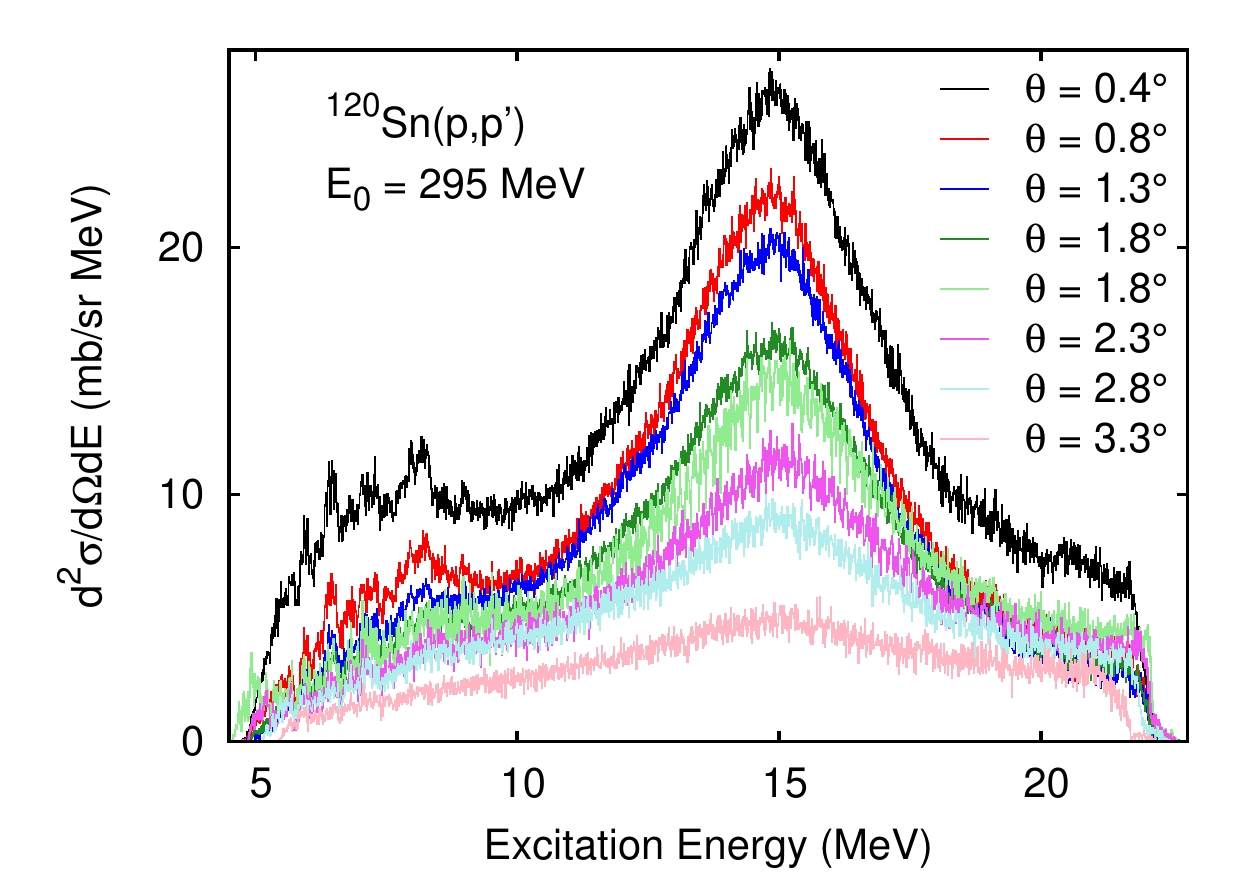}
\caption{\label{Analysis-120Sn-2008-CrossSection-AngleCuts} Experimental cross section of the $^{120}$Sn(p,p$^{\prime}$) reaction at $E_{\rm p} = 295$~MeV for different angle cuts. 
The top four spectra origin from the measurement with the Grand Raiden spectrometer angle set to  $0^\circ$, whereas the lower four were taken at $2.5^\circ$.
\label{fig:spectra}}
\end{figure}
Experimental cross sections of the $^{120}$Sn(p,p$^\prime$) reaction for various angle cuts are shown in Fig.~\ref{Analysis-120Sn-2008-CrossSection-AngleCuts}. 
A typical energy resolution of 30 keV (full width at half maximum, FWHM) was achieved.
The top four histograms correspond to the data with the Grand Raiden spectrometer set to $0^\circ$ and the lower four to data taken at $2.5^\circ$. 
The angular acceptance of the $0^\circ$ and $2.5^\circ$ setup overlap, so that for $\Theta=1.8^\circ$ two independent results can be shown. 
They agree well.

The dominance of relativistic Coulomb excitation under these kinematic conditions \cite{tam11,pol12} leads to prominent excitation of the GDR centered at about 14 MeV in the spectra for the most forward angles.
At lower excitation energies  pronounced structures are visible at $\Theta=0.4^\circ$ and $0.8^\circ$ which slowly disappear towards larger angles.
The angular dependence indicates a dipole character of the excited states. 
At smaller angles the spectra show a local minimum around 9 MeV which also vanishes for larger angles.

\section{Multipole decomposition}
\label{sec:mda}

A multipole decomposition analysis (MDA) of cross-section angular distributions was performed similar to the one described in Ref.~\cite{pol12} based on a least-square fit of the type 
\begin{equation}
 \frac{\textnormal{d}\sigma}{\textnormal{d}\Omega}(\Theta,E_{\rm x})_{\textnormal{exp}} = \sum_{J^\pi} a_{J^\pi}  \frac{\textnormal{d}\sigma}{\textnormal{d}\Omega}(\Theta,E_{\rm x},J^\pi)_{\textnormal{DWBA}},
\label{eq:fit}
\end{equation}
 where all coefficients $a_{J^\pi} > 0$.
Data were summed over bins of 200~keV and 400~keV below and above 11.5 MeV, respectively.
Theoretical angular distributions based on quasiparticle-phonon model (QPM) calculations for $^{120}$Sn \cite{oze14}  calculated with the code DWBA07 \cite{dwba07} were used as input. 
As demonstrated for the case of $^{208}$Pb \cite{pol12}, the low momentum transfers of the experiment permit a restriction of multipoles in Eq.(\ref{eq:fit}) to E1, M1 and E2.
Because of experimental problems during the data taking only limited use could be made of additional data taken at a spectrometer angle setting of $4^\circ$. 
The reduced number of data points compared to Ref.~\cite{pol12} required the dissection of the spectrum into three energy regions ($< 9.2$ MeV, $9.2 - 12.8$ MeV, $> 12.8$ MeV) with partly different additional constraints: \\
(i)
The contributions due to excitation of the isoscalar giant quadrupole resonance (GQR) were subtracted from all spectra \cite{kru14}.
The corresponding cross sections were calculated with DWBA07 using the isoscalar E2 strength distribution extracted in an ($\alpha,\alpha'$) experiment \cite{li10} and the theoretical GQR angular distribution taken from the QPM results. \\ 
(ii)
Only two theoretical E1 angular distributions of the transitions with the largest B(E1) values in each of the three energy regions were considered. \\
(iii)
The angular distribution of the spinflip M1 strength was described by a single curve corresponding to the transition with the largest strength. 
This is a good approximation for small-angle proton scattering \cite{hey10}.
At energies above 12.8 MeV the contributions from spin-flip M1 strength were excluded based on the properties of the spinflip M1 mode in medium-mass and heavy nuclei \cite{hey10}. \\ 
(iv)
In the excitation energy region between 9.2 and 12.8 MeV the E1 cross sections were determined by a least-square fit to photoabsorption data \cite{ful69,lep74,uts11} converted to (p,p$^\prime$) Coulomb excitation cross sections.
The photoabsorption cross sections were approximated as the sum of a Lorentzian with parameters from Ref.~\cite{ful69} describing the tail of the GDR and a polynomial of fifth order at lower excitation energies \cite{kru14}. \\
The $a_{J^\pi}$ coefficients were then determined by a $\chi^2$-weighted averaging over fits for all possible combinations. 

\begin{figure}[tbh]
\centering
\includegraphics[width=\columnwidth]{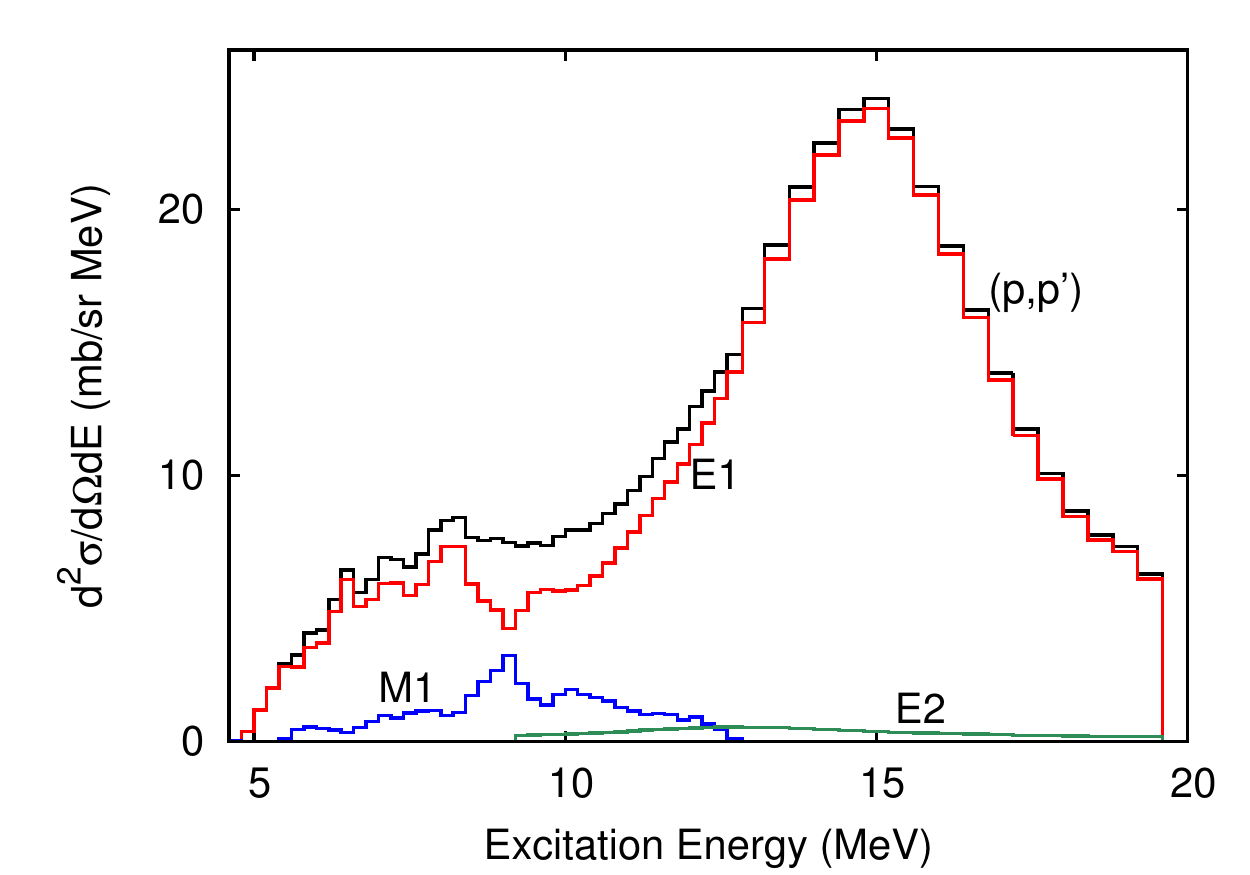}
\caption[Decomposition of the (p,p') spectrum]{Decomposition of the $^{120}$Sn(p,p$'$) spectrum at $E_{\rm p} = 295$ MeV and $\Theta = 0.4^\circ$ in terms of the multipoles E1, M1 and E2 with the MDA described in the text.\label{fig:Result-120Sn-MDA-FirstResult}}
\end{figure}

Figure \ref{fig:Result-120Sn-MDA-FirstResult} displays the resulting cross sections of E1, M1 and E2 multipoles for the $\Theta = 0.4^\circ$ spectrum.
Similar to the findings in $^{208}$Pb, E1 cross sections dominate at all excitation energies.
M1 cross sections contribute between 6 and 12 MeV with a maximum around 9 MeV.
The cross section contributions from the GQR determined with the above described procedure and representing about 100\% of the energy-weighted sum rule are very small.

\section{Low-energy E1 strength}
\label{ch:Comparison_with_NRF}

\begin{figure}[tbh!]
\centering
\includegraphics[width=\columnwidth]{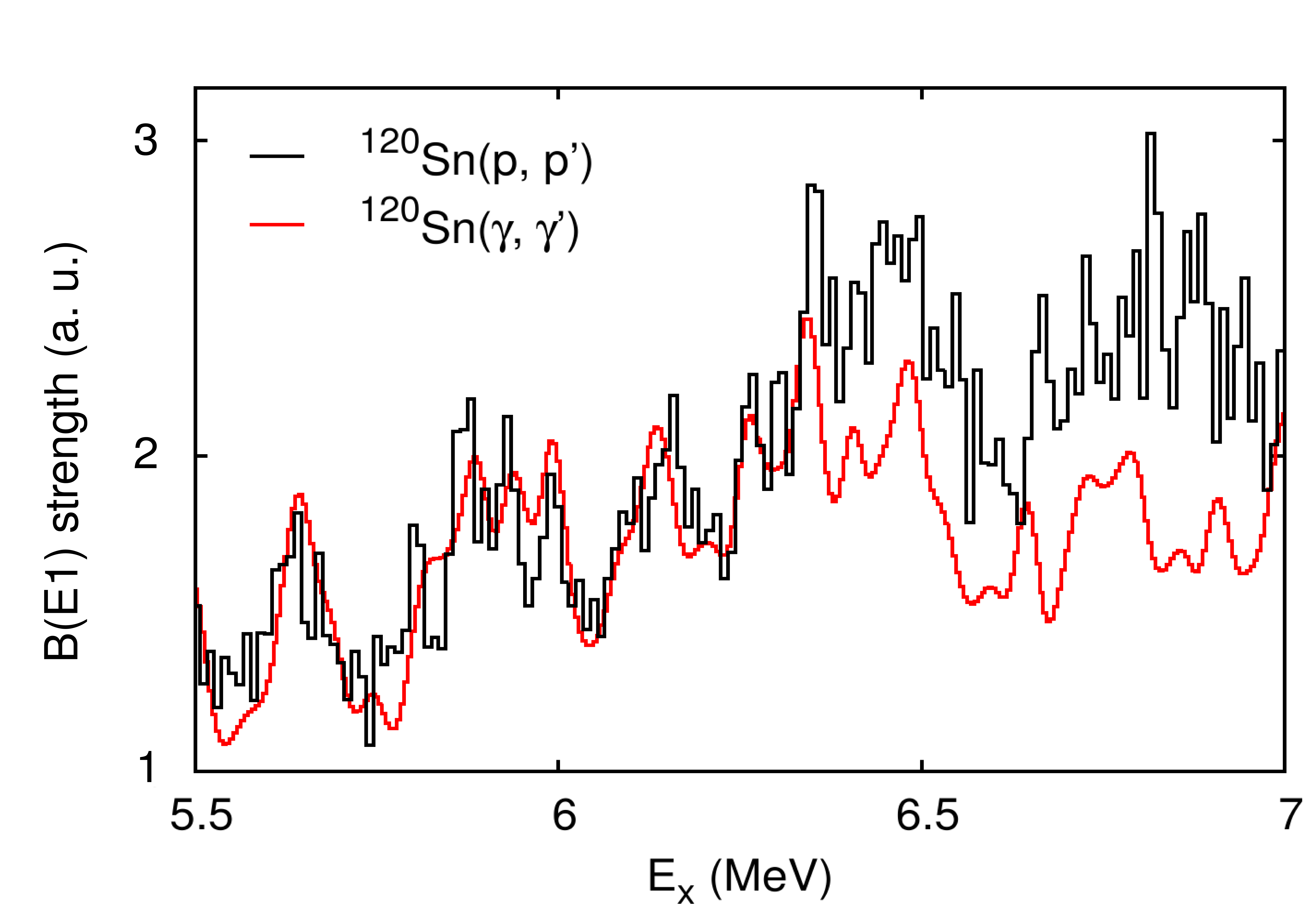}
\caption{Qualitative comparison of the E1 strength deduced from the $^{120}$Sn$(p,p')$ reaction at \mbox{$E_{\rm p} = 295$ MeV} and \mbox{$\Theta=0^\circ-0.5^\circ$} with the result from the $^{120}$Sn($\gamma$,$\gamma'$) reaction~\cite{oze14}. \label{fig:Results-GammaVsProtons}}
\end{figure}

The spectra of Fig.~\ref{fig:spectra} show considerable structure in the low-energy region. 
It is interesting to see whether these show correspondence to the E1 strength distribution deduced from the $^{120}$Sn($\gamma,\gamma'$) measurement \cite{oze14}.  
 A qualitative comparison of both experiments is shown in Fig.~\ref{fig:Results-GammaVsProtons}. 
For that purpose the background-subtracted ($\gamma$,$\gamma'$) spectrum with a resolution of better than 10 keV (FWHM) was folded with a Gaussian of width 30~keV (FWHM) to make it comparable to the proton scattering data. 
The (p,p$'$) spectrum was restricted to a very forward angle range $\Theta=0^\circ-0.5^\circ$ to enhance the E1 contribution.
Both spectra were normalized at the prominent peak around 5.6 MeV. 
Figure~\ref{fig:Results-GammaVsProtons} reveals good correspondence of the two experiments up to an energy of about 6.5 MeV. 
The smaller strength from ($\gamma,\gamma'$)  at higher energies indicates that either statistical decay of the exicted $1^-$ states to low-lying excited states becomes relevant and/or an increasing amount of fragmented strength falls below the sensitivity limit with increasing excitation energy.

\begin{figure}[t]
\centering
\includegraphics[width=\columnwidth]{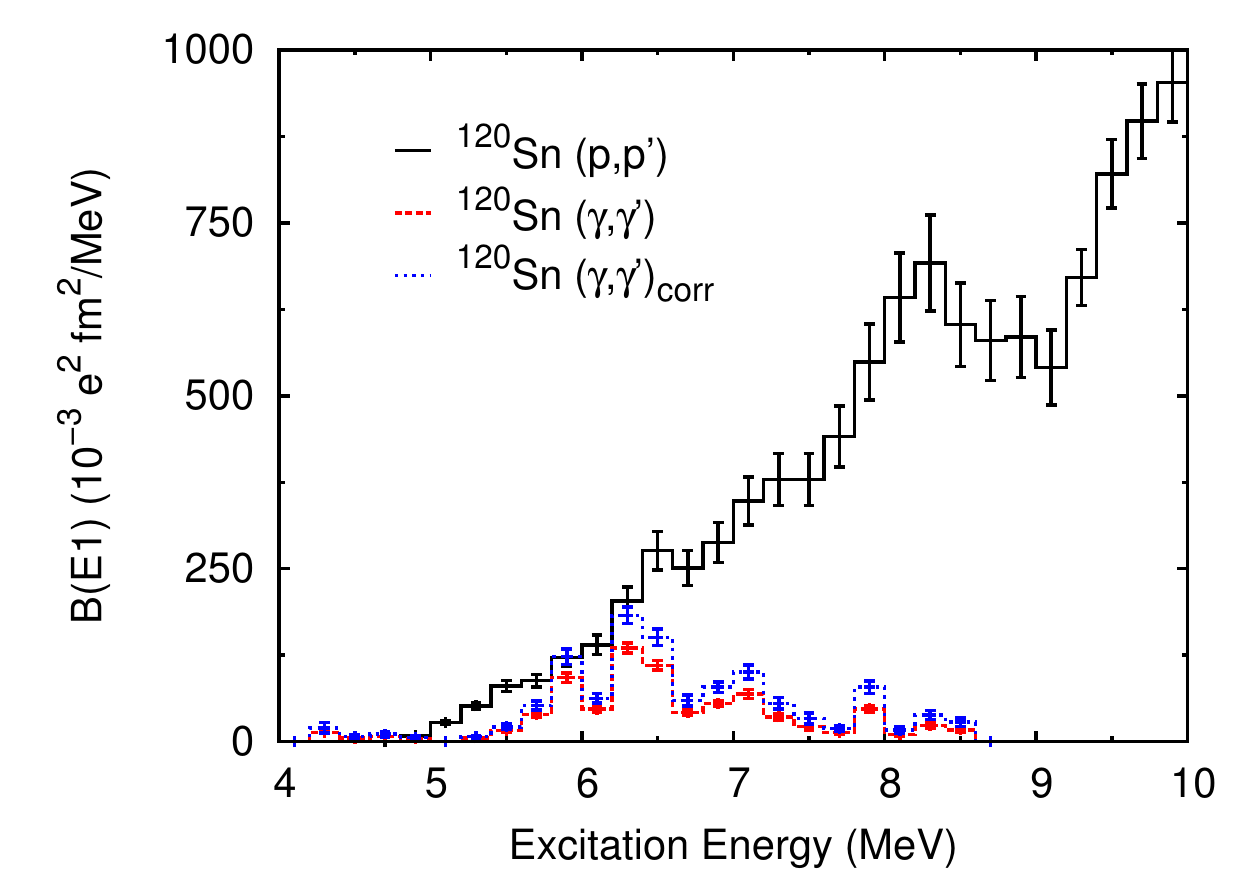}
\caption{\label{fig:Result-BE1-pp-gg}
B(E1) strength distribution of $^{120}$Sn in 200 keV bins from proton (present work) and photon scattering \cite{oze14}. 
$^{120}$Sn$(\gamma,\gamma')_\textnormal{corr}$ shows the strength corrected for branching ratios from a statistical model calculation~\cite{rus08} using the level density parameterization of Ref.~\cite{rau97}. 
}
\end{figure}

After the qualitative comparison of structures in the E1 strength distribution deduced from the (p,p$^\prime$) and $(\gamma,\gamma')$ experiments we now turn to a quantititave analysis. 
At very forward angles ($<$$1^\circ$) contributions from Coulomb-nuclear interference to the E1 cross sections are negligible \cite{pol12}.
The E1 strength distribution in $^{120}$Sn can be derived from the (p,p$'$) data in the semiclassical approximation \cite{ber88}.  
The B(E1) strength distributions given in 200~keV bins extracted from both experiments in the energy region up to 10 MeV are compared in Fig.~\ref{fig:Result-BE1-pp-gg}.
Note that the $(\gamma,\gamma^\prime)$ measurement is limited to energies below the neutron threshold ($S_n=9.1~$MeV). 
As pointed out already, above 6.3~MeV the (p,p$'$) results (black histogram) and the $(\gamma,\gamma')$ result (red circles) diverge. 
In the (p,p$'$) results a resonance-like structure at about 8.3 MeV is visible which is not seen in the $(\gamma,\gamma')$ data. 
Even in the energy region around 6 MeV, where good qualitative agreement between the two experiments is observed, the absolute B(E1) strength from the present work is about $20 - 40$\% larger.

The blue squares show a strength distribution from the $(\gamma,\gamma')$ data after correction for the unknown ground state branching ratios by a statistical model calculation. 
The method is described in Ref.~\cite{rus08}; for details of the application to $^{120}$Sn see Ref.~\cite{oze14}. 
The result shown in Fig.~\ref{fig:Result-BE1-pp-gg} uses the level density parameterization from Ref.~\cite{rau97}, but the dependence on the choice of the level density model is weak \cite{oze14}.
Inclusion of the statistical model correction brings both results in fair agreement in the energy region around 6 MeV.
Remaining differences may be related to the presence of unresolved strength in the $(\gamma,\gamma')$ data which was shown to be non-neglible \cite{oze14}. 
However, the sizable differences at higher excitation energies cannot be explained.
In general, while the increase of the E1 strength due to the statistical model corrections can be large in more deformed nuclei \cite{rus08,mas14}, in the semimagic nucleus $^{120}$Sn it does not exceed 40\% and thus cannot explain the orders-of-magnitude difference observed at excitation energies $> 7$ MeV.

The total exhaustion of the E1 EWSR up to 9.0 MeV corresponds to 2.3(2)\%, about twice the PDR strength in $^{208}$Pb \cite{pol12}.
In passing we note that the empirical relation for the B(E1) strength integrated over the excitation energy range $6-8$ MeV discussed in Ref.~\cite{mas14} gives a too large value for the present case. 
While Eq.~(3) in \cite{mas14} predicts 0.93 e$^2$fm$^2$, the experimental strength amounts to 0.57(4) e$^2$fm$^2$.

\begin{figure}[tbh!]
\centering 
\includegraphics[width=\columnwidth]{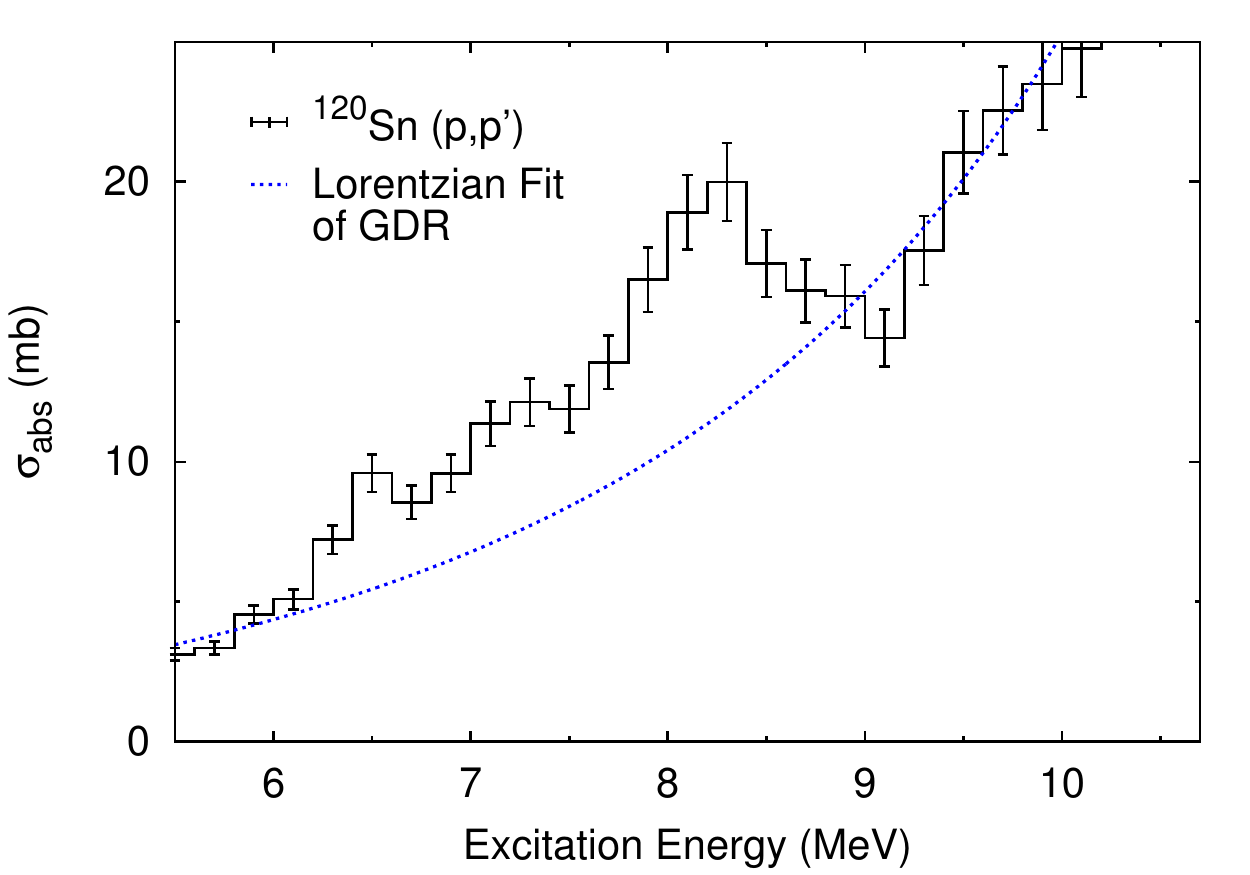}
\caption{\label{fig:Result-SigmaAbs-Lorentzian-PDR-2x1} 
Low energy photo-absorption cross section from the present experiment compared to a Lorentzian fit of the GDR (dashed blue line) with the parameters given in the text. 
}
\end{figure}
It is an open question to what extent the E1 strength distribution in $^{120}$Sn below threshold can be interpreted as the low-energy tail of the GDR.
In Fig.~\ref{fig:Result-SigmaAbs-Lorentzian-PDR-2x1} the conversion to photoabsorption cross sections is shown. 
A Lorentzian fit (dashed blue line) with parameters $\sigma_{\rm max} = 252$ mb, $E_{\rm c} = 15.0$ MeV, $\Gamma = 4.2$ MeV matches the energy regions below 6 MeV and above 9 MeV quite well with extra strength in between, which corresponds to 0.67(7)\% of the EWSR. 
The Lorentzian parameters deduced from $(\gamma$,xn) experiments \cite{ful69,lep74} cannot be used here because they overshoot on the lower-energy side of the resonance due to an asymmetry of the photabsorption cross sections in $^{120}$Sn towards higher excitation energies.
The present parameterization is restricted to an energy range where an approximately symmetric resonance form is observed.   
At excitation energies below 5 MeV there is very little E1 strength in $^{120}$Sn \cite{oze14} and the Lorentzian extrapolation overestimates the photoabsorption cross sections. 
This is a well-known problem for magic and semimagic nuclei and other parameterizations might be more appropriate~\cite{kop90}.  

The above decomposition suggests possibly the existence of two classes of distinct $1^-$ states: One consists of a number of selected states, in the present case around 6 MeV, with large g.s.\ decay probability. 
However, the larger part of the B(E1) strength seems to come from states with non-negligible ground-state decay width $\Gamma_0$ but ground-state branching ratios decreasing with excitation energy \cite{sch13}.
A recent study of the decay pattern of the $^{94}$Mo$(\gamma,\gamma^\prime)$ reaction finds a resonance-like structure between 5.5 and 7.5 MeV deviating from otherwise statistical decay \cite{rom13} in support of such a picture.  
Alternatively, one would have to assume that the level density of $1^-$ states in $^{120}$Sn is significantly higher than any model prediction.
This is unlikely in view of the fair reproduction of experimental level densities in Sn isotopes derived from thermal neutron capture \cite{cap09} and from a fluctation analysis \cite{pol14} of the fine structure of the GDR \cite{ebe15}. 

The bump around 6.5 MeV may be considered as the 'true' PDR.
This is supported by investigations of the isospin structure using the $(\alpha,\alpha')$ reaction \cite{end10} (although performed for $^{124}$Sn it is reasonable to assume that the properties of such an experiment on $^{120}$Sn would be similar).
The nature of the pronounced peak around 8.3 MeV is presently unclear.
A possible interpretation as local concentration on the low-energy tail of the GDR is discussed in Ref.~\cite{oro98}.

Finally, we recognize similar findings in $(\gamma,\gamma^\prime)$ experiments near shell closures which take into account quasi-continuum contributions in the spectra. 
In $^{90}$Zr a similar exhaustion of the EWSR and a factor of about 2.5 between the total scattering cross sections including unresolved parts and the analysis of discrete transitions was observed \cite{sch08}.   
Comparable E1 strengths were also seen near $N = 82$ in $^{136}$Ba \cite{mas12} and $^{138}$Ba \cite{ton10}, but a significantly larger part is concentrated in resolved g.s.\ transitions..
This is somewhat surprising since the level densities should be similar to $^{120}$Sn. 

\section{Comparison with model calculations}
\label{sec:comp}

\begin{figure}
\centering
 \includegraphics[width=\columnwidth]{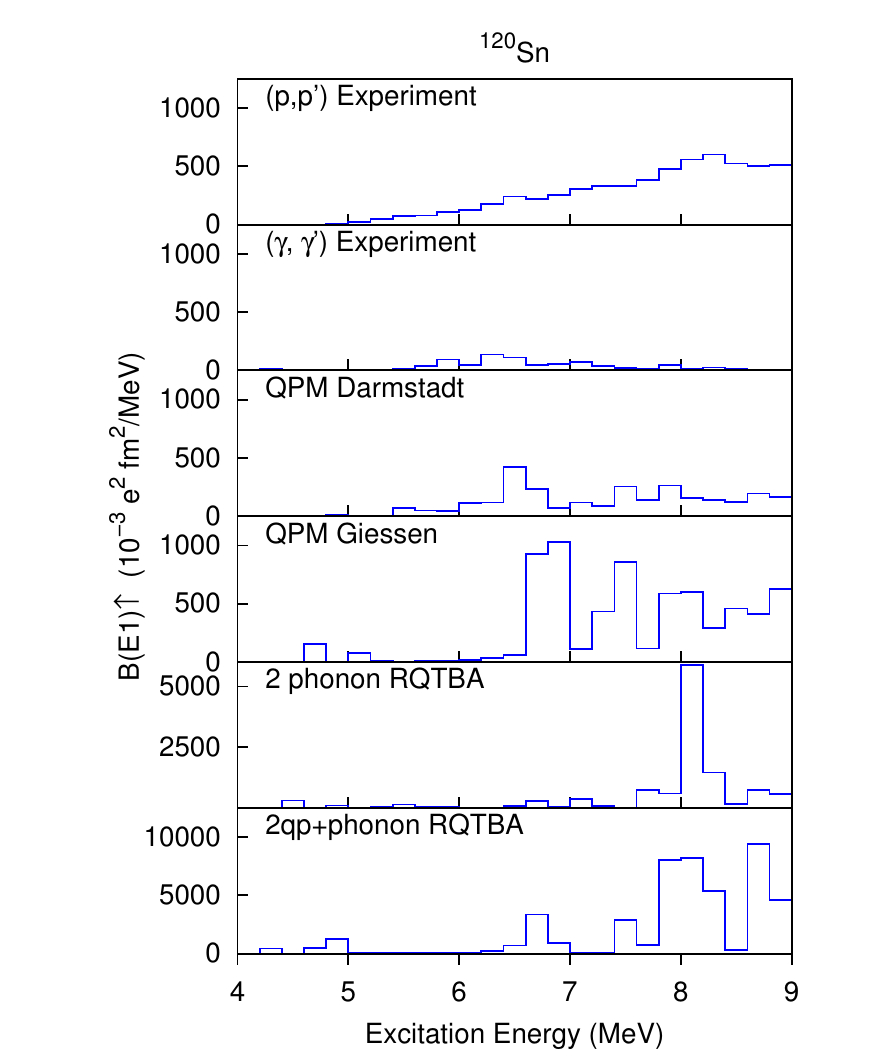} 
 \caption[B(E1) strength distribution for the (p,p') and ($\gamma,\gamma'$) data and microscopic models]{B(E1) strength distribution for the (p,p') and ($\gamma,\gamma'$) data and microscopic models in the energy range $E_{x}=4$\,--\,$9~$MeV in 200~keV bins. 
\label{fig:Results-Comparison-QPM-RQTBA-120Sn-BE1}}
\end{figure}

Theoretical predictions of the PDR in $^{120}$Sn show large variations.
Here we focus on a comparison with approaches including the coupling to complex states beyond the mean-field level, which can change low-energy E1 strength distributions considerably. 
Figure~\ref{fig:Results-Comparison-QPM-RQTBA-120Sn-BE1} presents the E1 strength distribution up to 9 MeV in $^{120}$Sn in energy bins of 200~keV. 
The plots include the two experimental results from the (p,p$^\prime$) and ($\gamma,\gamma^\prime$) data.
Two different quasi-particle phonon model (QPM) calculations in this work named QPM Darmstadt \cite{kru14} and QPM Giessen \cite{tso08} are presented. 
Both include coupling of 1-phonon (the RPA solutions) to 2- and 3-phonon states but use different ways to determine parameters of the underlying mean field and the residual interaction as described in Ref.~\cite{oze14}. 
Two calculations stem from the relativistic quasi-particle time blocking approximation (RQTBA) \cite{lit10}. 
They are based on a relativistic mean field approach but have different model spaces: 
The two quasi-particle phonon space (2qp+phonon) is built of quasi-particles from a relativistic mean field calculation which can couple to a phonon from the self-consistent renormalized quasi-particle RPA \cite{lit07}. 
In the two phonon space (2~phonon), all couplings between phonons are included \cite{lit13}. 
Further details are given in Ref.~\cite{oze14}.

The QPM Darmstadt result shows a peak of the E1 strength between 6 and 7~MeV, which roughly corresponds with the experimental results in this energy region. 
The result from QPM Giessen exhibits a similar peak but more strength at higher excitation energies.
The E1 strength predicted from the two RQTBA approaches differ from each other. 
The 2 phonon RQTBA result peaks around 8 MeV but predicts too little strength at lower excitation energies. 
The 2pq+phonon result is broadened compared to the 2 phonon result and the total  strength is four times larger. 
Both RQTBA calculations predict a strong rise of the strength above 7 MeV which conforms better with the distribution derived from the present data.

The summed B(E1) values for the energy region of $4-9$~MeV are given in Table~\ref{tab:summedbe1}. 
The results differ significantly. 
The summed B(E1) strength obtained by the present (p,p$^\prime$) experiment is about a factor of seven larger than the ($\gamma,\gamma'$) strength found in discrete transitions and still a factor of five larger after inclusion of corrections for unobserved branching ratios. 
After consideration of unresolved contributions deduced in Ref.~\cite{oze14} with a fluctuation analysis the present work finds more than three times E1 strength below threshold than the ($\gamma,\gamma'$)  data.
Concerning the total strength the QPM Giessen result is closest to the (p,p$^\prime$) experiment. 
The QPM Darmstadt result shows less strength, while both RQTBA calculations predict much higher strengths than seen experimentally.
\begin{table}[t]
\centering
\caption{ Experimental B(E1) strengths in $^{120}$Sn summed between 4 and 9 MeV and corresponding theoretical results from the calculations shown in Fig.~\ref{fig:Results-Comparison-QPM-RQTBA-120Sn-BE1}.}
\begin{tabular}{lcc}
\hline
& Ref. & $\Sigma B(E1)$ (e$^2$fm$^2$)   \\
\hline
$(p,p')$  & present  & 1.169(12) \\
$(\gamma,\gamma')$  & \cite{oze14} & 0.164(31)    \\
$(\gamma,\gamma')_{\rm corr}$  & \cite{oze14} & 0.228(43)    \\
$(\gamma,\gamma')_{\rm corr}$ + unresolved  & \cite{oze14} & 0.348(76)    \\
QPM Darmstadt  & \cite{kru14}  &  0.553 $\phantom{111}$    \\
QPM Giessen     & \cite{tso08} & 1.364  $\phantom{111}$     \\ 
2 phonon RQTBA   & \cite{lit13}  &  2.344  $\phantom{111}$ \\
2q+phonon RQTBA  & \cite{lit07}  & 9.494  $\phantom{111}$ \\
\hline 
\end{tabular}
\label{tab:summedbe1}
\end{table}

\section{Conclusions and outlook}

We have presented a measurement of the $^{120}$Sn (p,p$^\prime$) reaction at energies of a few hundred MeV and at extreme forward angles, where relativistic Coulomb excitation dominates the cross sections.
The method allows for the first time a consistent study of E1 strength in $^{120}$Sn below and above threshold in a single experiment.
The present results show a more than three times larger E1 strength below neutron threshold than derived from the ($\gamma,\gamma'$) experiment \cite{oze14} with a very different excitation energy distribution.
While the B(E1) strength distributions agree fairly well between 5.5 and 6.5 MeV, the present work finds much larger strengths at higher and lower excitation energies.
The latter is quite surpising, since one would expect from other studies at shell closures that the  ($\gamma,\gamma'$) reaction sees most of the strength at these low energies.
This point needs further experimental investigation.

Since low-energy E1 strength is a global phenomenon in nuclei with neutron excess one may expect comparably large effects for other cases.
Thus, all attempts to study systematics of the PDR based solely on strengths deduced from g.s.\ transitions in ($\gamma,\gamma'$) experiments should be viewed with some care. 
Similar conclusions were drawn from the analysis of ($\gamma,\gamma'$) experiments including a quasi-continuum part \cite{ton10,sch08,mas12} and in recent studies of the decay pattern \cite{sch13,rom13}. 
On the other hand, very good correspondence of the results from both probes was observed in $^{208}$Pb \cite{pol12}.

Clearly, a systematic study of complete E1 strength distributions with the (p,p$'$) reaction in nuclei at different shell closures but also extending to more deformed nuclei is called for. 
An improved understanding of the structure phenomena indicated by the present results may be achieved with coincidence studies of  $(\alpha,\alpha'\gamma)$ and (p,p$'\gamma$) reactions envisaged at RCNP in the frame of the CAGRA collaboration \cite{CAGRA}, by investigations of the $(\gamma,\gamma'\gamma'')$ reaction at HIGS \cite{isa13} and with the NEPTUN tagger \cite{sav10} at the S-DALINAC in Darmstadt, or by application of the self-absorption technique \cite{rom15}.    

\section*{Acknowledgements}

We thank the accelerator crew at RCNP for providing excellent beams. 
Discussions with N.~Pietralla and D.~Savran are gratefully acknowledged.
This work has been supported by the DFG (contracts SFB 634 and NE 679/3-1) and by the JSPS (grant No. 25105509).
N.T.K.\ acknowledges support from NAFOSTED of Vietnam under grant 103.01-2011.17.

\end{document}